\documentclass[aps,pre,twocolumn]{revtex4-1}
\usepackage[dvips]{graphicx,rotating}
\usepackage{amssymb}
\usepackage{amsmath}
\usepackage{epsfig}
\usepackage{bm}
\usepackage{amsfonts,float}
\usepackage{hyperref}

\usepackage{graphicx}
\usepackage[usenames,dvipsnames]{color}

\newcommand{\beq}{\begin{equation}}
\newcommand{\eeq}{\end{equation}}
\newcommand{\eps}{\epsilon}
\newcommand{\mS}{m_{\rm S}}
\newcommand{\mU}{m_{\rm U}}
\newcommand{\ee}{{\rm e}}

\begin{document}

\title{Large deviations in models of growing clusters with symmetry-breaking transitions}

\author{Robert L. Jack}
\affiliation{Department of Applied Mathematics and Theoretical Physics, University of Cambridge, Wilberforce Road, Cambridge CB3 0WA, United Kingdom}
\affiliation{Department of Chemistry, University of Cambridge, Lensfield Road, Cambridge CB2 1EW, UK}

\begin{abstract}
We analyse large deviations of the magnetisation in two models of growing clusters.  
The models have symmetry-breaking transitions, so the typical magnetisation of a growing cluster may be either positive or negative, with equal probability.  
For large clusters, the magnetisation obeys a large deviation principle.  We show that the corresponding rate function is zero for values of the magnetisation that are intermediate between the two steady state values, which means that fluctuations with these values of the magnetisation are much less unlikely than previously thought.  We show that their probabilities decay as power laws in the cluster size, instead of the exponential scaling that would be expected from the large deviation principle.  We discuss how this observation is related to dynamical phase coexistence phenomena.  We also comment on the typical size of magnetisation fluctuations.
\end{abstract}

\maketitle

\section{Introduction}
\label{sec:introduction}

Large deviation theory is the mathematical framework for analysis of rare events~\cite{denH-book}.  In recent years, this theory has been applied to a wide range of physical systems, in order to study dynamical fluctuations~\cite{Lebowitz1999,Bodineau2004,Bertini2015,Mehl2008,Derrida2007,Lecomte2007,Tailleur2007,Garrahan2007,Hurtado2014,kggw18}.  
In non-equilibrium systems, large-deviation theory can be used to analyse fluctuations of currents and of the entropy production~\cite{Lebowitz1999,Bodineau2004,Mehl2008,Bertini2015,Derrida2007}, with implications for fluctuation theorems~\cite{Lebowitz1999,Harris2007} and thermodynamic uncertainty principles~\cite{Gingrich2016}.
In systems that exhibit metastability, including glasses~\cite{Garrahan2007,Garrahan2009,Hedges2009} and biomolecules~\cite{Weber2014protein}, large deviation theory can be used to analyse deviations from ergodic behaviour, generating new insights into long-lived metastable states.  

To illustrate
the simplest case,  consider a system with (time-dependent) state $x(t)$, which follows some stochastic dynamics.
Select an observable quantity $F(x)$, and construct its time average as
\beq
\overline{F}(\tau) = \frac{1}{\tau} \int_0^\tau F(x(t)) \mathrm{d}t  \; .
\label{equ:fbar}
\eeq
This quantity is a random variable.  A simple statement of ergodicity is that $\overline{F}$ should converge to a corresponding ensemble average, in the limit where $\tau\to\infty$.  Large deviation theory provides a framework in which this convergence can be analysed~\cite{Derrida2007,Lecomte2007,Maes2008,Garrahan2009}.  Definitions will be given below, here we give an outline of the physical picture.  
In simple cases,  the probability density for $\overline{F}$ scales for $\tau\to\infty$ as
\beq
p\!\left( \overline{F} | \tau\right) \simeq \exp\left[ -\tau I\!\left(\overline{F}\right) \right] \; .
\label{equ:ldp-intro}
\eeq
This type of scaling relationship is called a large deviation principle (LDP) and
$I$ is called the rate function~\cite{denH-book,Touchette2009}.  For physical systems that are finite and ergodic, one expects that the rate function is (strictly) convex, with a single zero when $\overline{F}$ is equal to the ensemble average.

However, there is a wide range of physical systems where this simple picture breaks down.  For example, in systems with long-lived metastable states, it is natural for the rate function to develop singularities~\cite{Garrahan2007,Jack2010rom}, which appear as the system size tends to infinity.  Systems with long-ranged memory may also behave non-ergodically~\cite{Harris2009,Harris2015,Harris2015-peak-end}, in which case the rate function may not be convex  

In this work, we consider
systems where clusters of particles grow, as a function of  time~\cite{Klymko2017,Morris2014}.  
The clusters contain two kinds of particles, which can be distinguished either by their spins (up or down), or their colors (red or blue).  Using the language of spins, one defines the (extensive) magnetisation, which is the difference between the number of spin-up particles and the number of spin-down particles.  We focus here on models that are symmetric under interchange of spin-up and spin-down.  In this case, Klymko, Garrahan, and Whitelam~\cite{Klymko2017} showed that these systems exhibit a range of interesting behaviour, including symmetry-breaking transitions.  That is, depending on the model parameters, the (intensive) magnetisation can converge to zero at large times, or it can converge to a non-zero value~\cite{Klymko2017}.  In the latter case, positive and negative values are equally likely, but typical trajectories of the model involve spontaneous breaking of the up-down symmetry.  More generally, these models are also members of a general class of (Polya) urn problems~\cite{Pemantle2007,Hill1980} that have been studied in mathematics; they can also be reformulated as random walks with memory, similar to the elephant random walk~\cite{Trimper2004} (see also~\cite{Usatenko2003,Hod2004,Harris2015-peak-end,Baur2016}).

Recently~\cite{kggw18}, Klymko, Geissler, Garrahan and Whitelam (KGGW) analysed LDPs similar to (\ref{equ:ldp-intro}), 
in these growth models.  Since the clusters are always growing with time, the models are not ergodic.  This means that one cannot apply standard methods from~\cite{Derrida2007,Lecomte2007,Chetrite2015} in order to establish LDPs similar to (\ref{equ:ldp-intro}).  
Nevertheless, KGGW showed that an LDP holds at large time, where the observable analogous to $\overline{F}$ is the magnetisation.

For parameters where the symmetry is spontaneously broken, Ref.~\cite{kggw18} also predicted two unusual properties of the rate function.  First, there are some values of the observable for which the rate function is concave (the second derivative is negative).  Second, in cases where the steady states spontaneously break the symmetry, they predicted that the rate function should have local minima at unstable fixed points of the dynamics.  From (\ref{equ:ldp-intro}), it follows that these unstable points should be associated with local maxima in the probability, although these maxima were not observed in~\cite{kggw18}.  These predictions were based on a method that provides upper bounds on the rate function~\cite{kggw18,Whitelam2018sampling}.   This is achieved by modifying the dynamics of the system of interest, in order to make the rare events of interest more likely.  A recent preprint~\cite{Jacobson-arxiv} discusses this method in more detail, and proposes a method for calculation of the difference between the upper bounds and the true rate functions.

In this work, we examine the LDPs of KGGW in more detail.  We find that while their upper bounds on the rate function are valid, these bounds are not quantitative estimates of the rate functions themselves.  
In particular, we show that the second derivative of the rate function may be zero, but it is never negative.    (The rate function is not strictly convex, but neither is it concave.)  This also explains why the probability distribution of the magnetisation does not have local maxima at unstable fixed points of the dynamics.

Our analysis also reveals additional interesting and unusual behaviour in these systems.  In cases where the symmetry is spontaneously broken, we find that the probability distribution of the magnetisation behaves similarly to (\ref{equ:ldp-intro}) for some values of the magnetisation, while for some other values it behaves as a power law 
$p\!\left( \overline{F} |\tau \right) \simeq \tau^{-\alpha}$.  We discuss the physical interpretation of this unusual scaling -- it is associated with the fact that very large fluctuations can be observed for trajectories that diverge slowly from the unstable fixed point of the dynamics (a similar phenomenon has been discussed before~\cite{Franchini2017} in Polya urn models).   In the light of these results, we discuss what general insights are available, including the strengths and weaknesses of numerical strategies for analysis of LDPs.  In particular, we emphasise that while upper bounds on rate functions are useful, establishing quantitatively accurate bounds is likely to require a detailed understanding of the system of interest~\cite{Jack2014-east,Jack2015b,Jack2010}, or a very flexible variational ansatz~\cite{banuls2019-arxiv}.


In this paper, Sec.~\ref{sec:models} describes the models and the main methods of~\cite{Klymko2017,kggw18}; this includes a model of irreversible cluster growth, and one where the cluster grows reversibly, so that the number of particles can both increase or decrease.  Sec.~\ref{sec:irrev-results} has results for the irreversible process, and Sec.~\ref{sec:rev-results} has results for the reversible process.  In Sec.~\ref{sec:conc} we discuss what general conclusions can be drawn from our results.

\section{Models and methods}
\label{sec:models}

This Section defines the models considered in this work, which were introduced by Klymko, Garrahan and Whitelam~\cite{Klymko2017}.  
We also describe the method used by KGGW to obtain bounds on the probabilities of rare events in these models~\cite{kggw18}.  The definitions of the models and methods are equivalent to those of~\cite{kggw18}.  We also discuss the relationships between those works and other models and methods from the literature.

\subsection{Irreversible model of growth}

In the irreversible model of growth defined in~\cite{Klymko2017}, clusters grow irreversibly (see also~\cite{Morris2014}).  
KGGW consider trajectories of this model that have a fixed number of events~\cite{kggw18}, which amounts to considering a stochastic process in discrete time.  Let $s_k=\pm1$ be the spin of the particle that is added on step $k$.  The (intensive) magnetisation just after this step is
\beq
m_k = \frac{1}{k}  \sum_{i=1}^k s_i \, .
\label{equ:def-m}
\eeq
The initial condition is that $s_1=\pm 1$, each with probability $1/2$.
The dynamical rule is that 
\beq
s_{k+1} = 
\begin{cases} 
+1, &  \hbox{with prob } (1+\exp(-2Jm_k))^{-1} \; ,
\\
 - 1, &  \hbox{with prob } (1+\exp(2Jm_k))^{-1} \; .
\end{cases} 
\label{equ:Mk+1}
\eeq
The parameter $J>0$ describes a ferromagnetic interaction among the spins.  We consider trajectories of this discrete-time model with $K$ steps in total: this set of trajectories is identical to the \emph{constant-event-number ensemble} of KGGW~\cite{kggw18}, where the number of events is $K$.


 There are several possible ways to analyse this growth process.  It can be interpreted as a Markov process for $m_k$, in which the transition probabilities depend explicitly on the step $k$.  (The process is not stationary.)   This model can be mapped onto a (generalised) Polya urn problem~\cite{Pemantle2007,Hill1980} by interpreting the number of up/down spins in the growing cluster as the number of red/black balls in a container (similar to~\cite{Klymko2017}).
 General results for large deviations in urn models have been obtained by Franchini~\cite{Franchini2017}: the results presented here are consistent with that work.
 Alternatively, the system can be viewed as a non-Markovian process for $s_k$, in which the transition probabilities depend on the history, via $m_k$~\cite{Whitelam2018slow,Whitelam2018sampling}.  This formulation is related to the elephant random walk~\cite{Trimper2004}, where $M_k = km_k$ is viewed as the position of a particle, which obeys a dynamical rule similar to (\ref{equ:Mk+1}) except that the probabilities of $s_{k+1}=\pm1$ are linear in $m_k$.
 Within this non-Markovian formulation, the irreversible growth model falls in the broad class considered by Harris and Touchette~\cite{Harris2009,Harris2015}.  They derived several general results for large deviations in models within this class.
 The results presented here are consistent with their theory.

The large-$k$ behaviour of this model depends on the value of $J$.  
For large $k$, the magnetisation $ m_k  $ converges to a limiting value.  It is convenient to refer to this as a steady state for the model, even though the cluster is constantly growing so the system is not stationary.
Given some $m_k$, the conditional average of $s_{k+1}$ is
$
\langle s_{k+1} \rangle_{m_k} = \tanh Jm_k .
$
For large $k$ one finds
\beq
\langle m_{k+1} \rangle_{m_k} \simeq m_k  + \frac{\tanh(J m_k) -  m_k}{k} \, .
\label{equ:eom-m}
\eeq
This means that steady state values of the magnetisation must solve $m = \tanh Jm$~\cite{Klymko2017}.  This equation is familiar from mean-field models of ferromagnets: there is only one solution if $J\leq 1$ but for $J>1$ there are three solutions.  In this case, the steady states have $m=\pm \mS$ where $\mS$ is the spontaneous magnetisation.   The point $m=0$ is an unstable fixed point of (\ref{equ:eom-m}) and is not a steady state of the growth model.
This symmetry-breaking transition is similar to transitions in the non-Markovian random walk models of~\cite{Harris2015-peak-end}, see in particular their (so-called) artificial model.

We consider trajectories with a total of $K$ steps, and we discuss an LDP that applies for large $K$.
The analogue of the time-average in (\ref{equ:fbar}) is $m_K$, as defined in (\ref{equ:def-m}). 
The LDP discussed by KGGW is similar to (\ref{equ:ldp-intro}): as $K\to\infty$ one has
\beq
p(m_K) \simeq \exp\left[ -K I(m_K) \right]
\label{equ:ldp-mk}
\eeq
where $I$ is the rate function.  This is an LDP with speed $K$~\cite{denH-book,Touchette2009}.
We assume that $I$ is continuous, which is consistent with our results and those of KGGW.  
Define
\beq
{\cal I}_K(m,\eps) = -\frac{1}{K} \log \mathrm{Prob}\!\left( |m_K-m| \leq \eps \right) \; .
\label{equ:def-calI}
\eeq
where $\mathrm{Prob}(\ldots)$ on the right hand side denotes the probability that $m_K$ is within $\epsilon$ of some value $m$.
From the theory of LDPs~\cite{denH-book}, one has 
\beq
\lim_{K\to\infty} {\cal I}_K(m,\eps) = \inf_{x\in [m-\eps,m+\eps]} I(x) \; .
\label{equ:rate-interval}
\eeq  
Hence, analysis of ${\cal I}$ provides information about the rate function.  In particular, the assumed continuity of $I$ means that
$I(m) = \lim_{\eps\to0} \lim_{K\to\infty} {\cal I}_K(m,\eps)$.  

\subsection{Reversible model of growth}
\label{sec:rev-model}

The reversible model considered by KGGW was introduced in~\cite{Klymko2017}.  It is similar to the irreversible one, except that particles may leave the cluster as well as being added to it.  (Note however, this is {not} a reversible Markov chain in the mathematical sense.)  The number of particles in the cluster after step $k$ is $N_k$ and its (extensive) magnetisation is $M_k$.   Also $m_k=M_k/N_k$ is the intensive magnetisation.  There are several LDPs that could be considered, including the joint probability distribution for $M_K/K$ and and $N_K/K$, which was analysed by KGGW~\cite{kggw18}.  Here we consider the behaviour of $m_k$, for which we expect that
\beq
p(m_K) \simeq \exp[ -K I(m_K) ] \; .
\label{equ:ldp-rev}
\eeq
This is analogous to (\ref{equ:ldp-mk}), although the functional form of $I$ will be different.

The dynamical rule for the model may be formulated in  terms of increments $\Delta M_k = M_{k+1} - M_{k}$ and $\Delta N_k = N_{k+1} - N_{k}$.  
The physical idea~\cite{Klymko2017} is that particles are added with rates that do not depend on their spin, but their unbinding rate is suppressed if they are aligned with the magnetisation of the cluster. Specifically
\beq
(\Delta N_k,\Delta M_k) = 
\begin{cases} 
(+1,+1), &  \hbox{w/prob } (c/z_k)
\\
(+1,-1), &  \hbox{w/prob } (c/z_k)
\\
(-1,+1), &  \hbox{w/prob } {\rm e}^{Jm_k} (1-m_k) / z_k
\\
(-1,-1), &  \hbox{w/prob } {\rm e}^{-Jm_k} (1+m_k)/ z_k
\end{cases}
\label{equ:Mk+1-rev}
\eeq
where 
\beq
z_k = 2(c + \cosh Jm_k - m \sinh Jm_k)
\label{equ:zz}
\eeq
is a normalisation constant.  Similar to the irreversible case, the set of trajectories of this discrete-time model is equivalent to the constant-event-number ensemble of the reversible growth model of~\cite{kggw18}.

Like the irreversible case, this model can also be mapped to generalised Polya urn problem, where particles can be removed from the container, as well 
as added.  The model can also be interpreted as non-Markovian random walk in two dimensions, where the position is $(M_k,N_k)$.  This would be a two-dimensional generalisation of the elephant random 
walk~\cite{Trimper2004}.

The  steady states of this model are described in~\cite{Klymko2017}. There are symmetry-breaking transitions, similar to the irreversible model, but with some additional complexity.  We give a brief summary of the relevant behaviour in Section~\ref{sec:rev-results}, which also includes results for this model.

\subsection{Bounds on probabilities via optimal control theory}
\label{sec:control}

One method for analysing large deviations is to establish upper bounds on the rate function, by considering processes where the rare events of interest become typical.  
This idea is common in the mathematical theory of large deviations~\cite{denH-book,dupuis-book}, its application in physics is reviewed in~\cite{Chetrite2015var}, which also discusses its connection with optimal control theory~\cite{Fleming85}.
This connection provides us with a useful terminology: we consider \emph{control forces} which are added to the system, in order to enhance the probability of rare events (for a physical example, see~\cite{Nemoto2019}).

In mathematical studies, one usually aims to prove bounds on rate functions.  In KGGW~\cite{kggw18}, bounds were evaluated numerically, by direct simulation of the growth model, see also~\cite{Whitelam2018sampling,Jacobson-arxiv}.  (This approach may be contrasted with other numerical methods~\cite{Nemoto2014,Nemoto2016,Nemoto2017first,Ray2018,Ferre2018,Nemoto2019}, which use control forces to aid the computation of rate functions or other large-deviation properties, instead of computing bounds.)

We outline the derivation of the relevant bound, following~\cite{kggw18,Whitelam2018sampling}.  The method is very general, we focus here on the irreversible growth model.
Let $p_k(s|m)$ be the probability that the $(k+1)$th particle has spin $s$, given that the magnetisation just after step $k$ is $m$.  The probability of a trajectory $\bm{m}=(m_1,m_2,\dots,m_K)$ for the original process is $P(\bm{m})=\prod_{k=1}^K p_k(s_k|m_{k-1})$, with $s_k=m_{k+1}-m_{k}$.  
In the irreversible model, (\ref{equ:Mk+1}) shows that $p_k(s|m) = [1+\exp(-2Jm s)]^{-1}$, for $k\geq 2$.  This is independent of $k$ because the update rule does not depend on $k$, except through the value of $m_k$.  The initial condition is specified by taking $p_1(s|m)=(1/2)$ (for $s=\pm1$).  

Now introduce a \emph{controlled system} in which the corresponding transition probabilities are $p^{\rm con}_k(s_{k+1}|m_{k})$.  In~\cite{kggw18,Whitelam2018sampling,Jacobson-arxiv}, this process is called the reference system.  
The average of some observable quantity $F$ in the controlled process is denoted by $\langle F \rangle_{\rm con}$, and the corresponding average in the original system is $\langle F \rangle$.  By considering the probabilities of individual trajectories, one sees that
\beq
\langle F \rangle = { \langle F \exp[-{\cal A}(\bm{m})] \rangle_{\rm con} } 
\label{equ:Fcon}
\eeq
where the action $\cal A$ is
\beq
{\cal A}(\bm{m}) = \sum_{k=1}^K \log \frac{  p^{\rm con}_k(s_k|m_{k-1}) }{ p_k(s_k|m_{k-1}) } \; .
\label{equ:def-A}
\eeq

We define an indicator function $\chi_\eps(x)$ that is equal to unity if $|x|<\eps$, and zero otherwise.  Then use (\ref{equ:def-calI}) to write ${\cal I}_K(m,\eps)  = -K^{-1} \log \langle \chi_\eps(m_K-m) \rangle$.  Using (\ref{equ:Fcon}) to express the right hand side in terms of averages with respect to the controlled process, and noting that ${\rm e}^x$ is convex, one obtains by Jensen's inequality~\cite{kggw18,Whitelam2018sampling} that
\beq
{\cal I}_K(m,\eps) \leq {\cal H}_K(m,\eps)
\label{equ:IK-bound}
\eeq 
with
\begin{multline}
{\cal H}_K(m,\eps) = -\frac{1}{K} \log \langle \chi_\eps(m_K-m) \rangle_{\rm con} 
\\ + 
\frac{ \langle \chi_\eps(m_K-m) \cdot {\cal A}[\bm{m}] /K \rangle_{\rm con} }{\langle  \chi_\eps(m_K-m) \rangle_{\rm con}} \; .
\label{equ:def-H}
\end{multline}
The first term on the right hand side of (\ref{equ:def-H}) is the log probability of the event that $|m_K-m|\leq \eps$, in the controlled process.  The second term is a conditional average of the action, which is obtained by averaging over the trajectories that realise this event.  If this event of interest is typical under the controlled dynamics, it is simple to evaluate ${\cal H}$ and hence to obtain an upper bound on ${\cal I}$.  Obtaining accurate bounds (with ${\cal H}\approx {\cal I}$) requires a good choice of the controlled process.

For ergodic Markov processes in the classes considered by~\cite{Lecomte2007,Jack2010,Chetrite2015} (which include irreducible finite-state Markov chains and a large class of stochastic differential equations), the search for suitable controlled processes is somewhat simplified.  In these cases, it can be shown~\cite{Maes2008,Jack2010,Chetrite2015} that for $K\to\infty$ one may achieve equality in (\ref{equ:IK-bound}) by using a controlled process where the transition probabilities do not depend on the step $k$ (except possibly for $k=1$, which controls the initial condition).  
%
However, the situation for growth models is more complicated because achieving equality in (\ref{equ:IK-bound}) may require a controlled process whose transition rates depend explicitly on $k$.  The new results that we obtain here are obtained by considering this type of controlled process.%


\begin{figure}
\includegraphics[width=7cm]{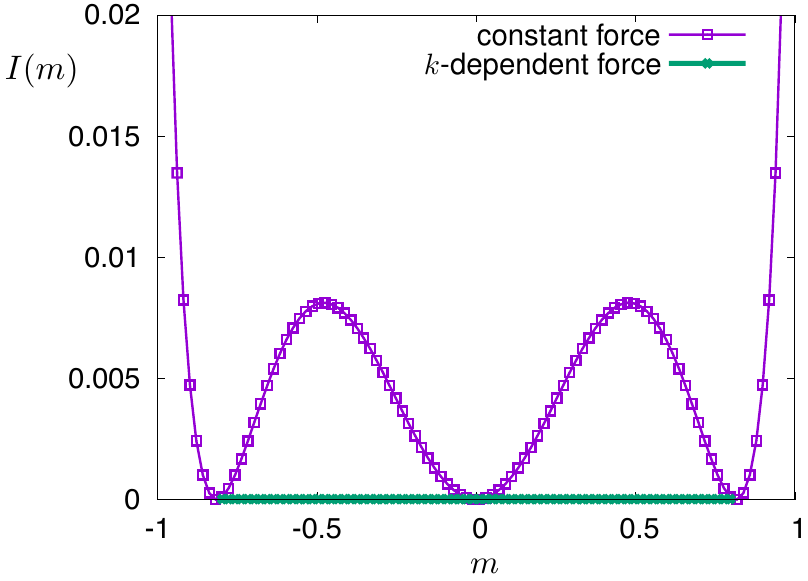} 
\caption{Upper bounds on the rate function for the LDP (\ref{equ:ldp-mk}) in the irreversible model, for the representative case $J=1.4$.  We show the bound of KGGW, which is obtained using control forces that are constant (independent of $k$).  This is compared with our new bound for $|m|<\mS$, which uses control forces that depend on $k$.  Our new result (with $k$-dependent forces) implies that $I(m)=0$ whenever $|m|<\mS$.}
\label{fig:rate-analytic}
\end{figure}

\section{Results : irreversible model}
\label{sec:irrev-results}

We analyse the LDP (\ref{equ:ldp-mk}) in the irreversible model.  KGGW derived bounds on the rate function $I$ that appears in (\ref{equ:ldp-mk}).  They argued that their bounds are equal to the rate function itself, because the controlled processes (reference processes) that they consider mimic the rare-event behaviour of the model.  We show here that their bounds are valid but they are not equal to the rate function.  Section~\ref{sec:results-ldp-irrev} focusses on the case $J>1$ in which the model spontaneously breaks up-down symmetry.  We show that the rate function in (\ref{equ:ldp-mk}) is zero throughout the range $|m|<\mS$  (see also Corollary 5 of~\cite{Franchini2017}).  
Fig.~\ref{fig:rate-analytic} summarises this result.  
After that, Sec.~\ref{sec:beyond-ldp}  discusses the behaviour of $p(m_K)$ for $m_K$ in this range: we show that this probability decays as a power law in $K$.  Then Sec.~\ref{sec:harris-regime} gives a brief discussion of the behaviour for $|m_K|\gtrsim\mS$, focussing on the behaviour close to the peaks of the probability distribution. 

\subsection{LDP with speed $K$}
\label{sec:results-ldp-irrev}

%
%
%

This Section establishes the new bound shown in Fig.~\ref{fig:rate-analytic}.  Before starting the calculation, we recall the physical interpretation of this rate function.
In many LDPs, zeros of the rate function $I(m)$ correspond to typical values of the observable $m$~\cite{Touchette2009}.   The situation shown in Fig.~\ref{fig:rate-analytic} is different, in that the typical values of $m$ are $\pm \mS$ but the rate function is zero throughout the range $|m|\leq\mS$.  This behaviour occurs when the probability density for $m$ decays to zero less fast than an exponential [for example $p(m_K)\sim K^{-\alpha}$] so that $p(m_K)\to 0$ at large $K$, but one still has ${\cal I}_K(m,\eps)\to0$, and hence $I(m)=0$ by (\ref{equ:rate-interval}). This behaviour may be somewhat unusual but it is fully consistent with the existence of a large deviation principle with speed $K$~\cite{denH-book}.  It is also similar to the behaviour of the free energy in thermodynamic systems close to phase coexistence, see Sec.~\ref{sec:conc}.

\subsubsection{Control forces that are independent of $k$}
\label{sec:control-kggw}

We first derive the bounds of KGGW, following their method, which uses (\ref{equ:IK-bound}).   We use a slightly different controlled system in which the extensive magnetisation 
$
M_k = k m_k
$
follows a biased random walk (see also~\cite{Whitelam2018sampling,Whitelam2018slow}).
The dynamical rule for this controlled system is
\beq
s_{k+1} = 
\begin{cases} 
+1, &  \hbox{with prob } (1+b)/2
\\
 - 1, &  \hbox{with prob } (1-b)/2
\end{cases}
\label{equ:Mk+1-control}
\eeq
where $b$ is a numerical parameter with $|b|\leq1$.
In this case, the mean and variance of $m_k$ under the controlled dynamics are
\beq
\langle m_k \rangle_{\rm con} = b, \qquad \langle(\Delta m_k)^2\rangle_{\rm con} = (1-b^2)/k  \; .
\label{equ:diff}
\eeq    
For large $k$, the variance tends to zero and $m_k$ becomes sharply peaked at $b$.  This allows calculation of ${\cal H}_K$ in (\ref{equ:def-H}), and hence bounds on ${\cal I}_K$.  For large $K$ then $ \langle \chi_\eps(m_K-b) \rangle_{\rm con}\to1$.
Also, from (\ref{equ:Mk+1-control}) one has $p^{\rm con}_k(s|m) = (1+sb)/2$, independent of $m$, so (\ref{equ:def-H}) yields
\begin{multline}
\langle {\cal A}(\bm{m}) /K \rangle_{\rm con} \simeq \frac{1+b}{2} \log \frac{(1+b)(1+\ee^{-2Jb})}{2}
\\ + \frac{1-b}{2} \log \frac{(1-b)(1+\ee^{2Jb})}{2}
\label{equ:A-homog}
\end{multline}
We used that the fraction of steps with $s_k=\pm1$ is $(1\pm b)/2$ and the contribution to the action for each such hop is $\log [(1\pm b)(1+\ee^{\mp2Jm})/2]$; also $m_k$ is sharply peaked at $b$, so the \emph{average} action for such a hop can be estimated as $\log [(1\pm b)(1+\ee^{\mp 2Jb})/2]$.  Since $m_k$ is sharply peaked, we note that this result for the action is somewhat insensitive to details of the controlled dynamics.  For example, if the rates in (\ref{equ:Mk+1-control}) depended on also $m_k$, the action would only be sensitive to the values of the rates at the mean value of $m_k$.  This insight is related to the temporal additivity principle of Harris and Touchette~\cite{Harris2009,Harris2015}.  In our context, it means that adding extra complexity to the controlled process (\ref{equ:Mk+1-control}) does not yield improved bounds on ${\cal I}$.

Using again that $m_K$ is sharply peaked, the conditional average of the action in (\ref{equ:def-H}) can be replaced by the simple average in (\ref{equ:A-homog}), and one obtains (after simplifying the right hand side of (\ref{equ:A-homog}) and setting $b=m$)
\begin{multline}
{\cal H}_K(m,\eps) \simeq \frac12 \log (1-m^2) + \log \cosh(Jm)
\\ + \frac{m}{2}  \log \frac{1+m}{1-m} - Jm^2 
\label{equ:HK-homog}
\end{multline}
as in~\cite{kggw18}.
This result is valid for large $K$.
It is easily checked that ${\cal H}_K(m,\eps)$ is non-negative for all $m$ (as it must be, since it is a bound on ${\cal I}$).  Using $2\tanh^{-1} m = \log (1+m)/(1-m)$, one also sees that ${\cal H}_K(m,\eps)=0$ if $m=\tanh(Jm)$. That means that ${\cal I}_K(m,\eps)=0$ if $m$ is fixed point of (\ref{equ:eom-m}). For $b\ll 1$ we also obtain
\beq
\langle {\cal A}(\bm{m}) /K \rangle_{\rm con} \simeq \frac{b^2(J-1)^2}{2} + {\cal O}(b^4)
\label{equ:A-small-b}
\eeq
which determines the action of trajectories with $m\approx 0$.  [Recall, we are considering $J>1$ so $m=0$ is an unstable fixed point of (\ref{equ:eom-m}).]

\begin{figure*}
\includegraphics[width=17.8cm]{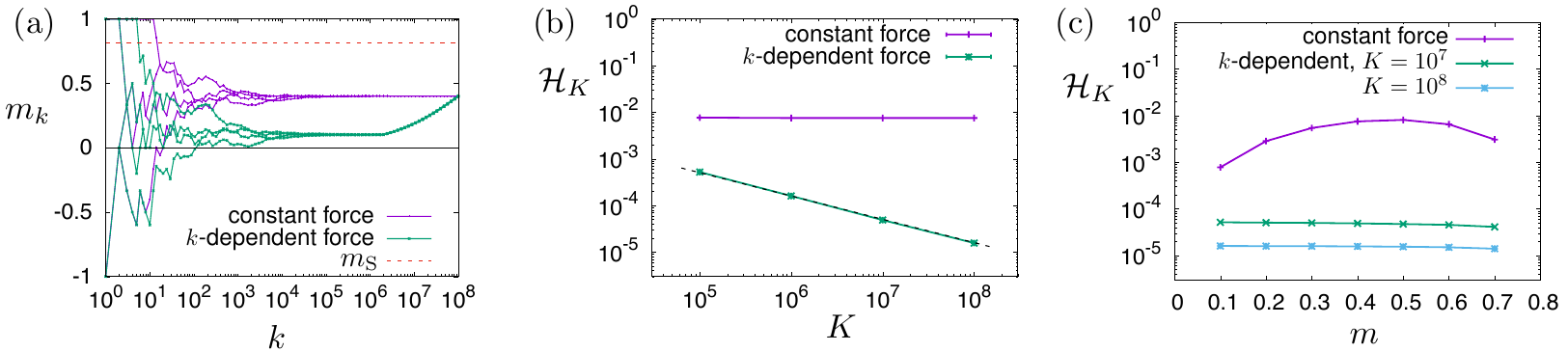}
\caption{Results for the irreversible growth model in the representative  case $J=1.4$, for which $\mS\approx 0.82$.  (a)~Trajectories of the controlled dynamics for $K=10^8$, which all achieve $m_K\approx0.4$.  For the $k$-dependent force then $(b,k^*) = (0.10,2.0\times 10^6)$.
(b)~Results for ${\cal H}_K$ using the same controlled dynamics, as a function of $K$.  (For the time-dependent forces, the values of $b,k^*$ are different for each $K$.) Error bars are comparable with symbol sizes.  The dashed line is the prediction of (\ref{equ:H-sqrtK}), without any fitting parameters.
(c)~The behaviour of ${\cal H}_K$ as a function of $m$. 
The data in (b,c) and the agreement with (\ref{equ:H-sqrtK}) indicate that $\lim_{K\to\infty} {\cal H}_K(m,\eps)=0$ throughout the range $|m|<\mS-\epsilon$, so $\lim_{K\to\infty} {\cal I}_K(m,\eps)=0$ within this range, by (\ref{equ:IK-bound}).}
\label{fig:rateFn}
\end{figure*}

At this fixed point, one sees that ${\cal I}_K(m,\eps)\simeq0$ (for large $K$).  This means that trajectories with $m_K=0$ have probabilities that do not decay exponentially in $K$.  The physical interpretation of this fact is that if the growing cluster contains a symmetric mixture of up and down spins, there is no force that acts to increase or decrease the magnetisation.  On the other hand, if $m$ is intermediate between $0$ and $\mS$, there is a force that drives the system towards the stable fixed point at $\mS$.  This is the intuition behind the LDP (\ref{equ:ldp-mk}): the probability to remain for a long time at a non-typical value of $m$ is suppressed exponentially in $K$, because of the forces in the model that tend to drive $m$ towards a typical value.  At $m=0$, the force is zero, so the probability to remain near this value is suppressed less strongly.

So far, all results are fully consistent with KGGW~\cite{kggw18}.  We now show that for $m<m_{\rm S}$, the true value of ${\cal I}_K$ is much less than ${\cal H}_K$ in  (\ref{equ:HK-homog}).   We will obtain improved bounds by taking
a controlled process in (\ref{equ:IK-bound}) in which the transition rates depend explicitly on the step $k$.

\subsubsection{Control forces that depend on $k$}

We take a controlled process that is a mixture of (\ref{equ:Mk+1-control}) and (\ref{equ:Mk+1}), as follows. 
We choose two parameters, which are the bias $b$ in (\ref{equ:Mk+1-control}) and a step $k^*$ at which the controlled dynamics changes its character. For the early part of the trajectory, which is $k\leq k^*$, the controlled system is an asymmetric random walk as in (\ref{equ:Mk+1-control});  for the later part ($k>k^*$) we revert to the original dynamics (\ref{equ:Mk+1}).  One sees that the action ${\cal A}$ in (\ref{equ:def-A}) has no contribution from the later part of the path.  Since smaller values of ${\cal A}$ lead to more accurate bounds, this is a desirable feature.  We restrict to $|b|<\mS$ which is sufficient for our purposes.

Typical trajectories of this controlled dynamics are illustrated in Fig.~\ref{fig:rateFn}(a).  If $b^2 k^*\gg 1$ then one sees from (\ref{equ:diff}) that the distribution of $m_{k^*}$ is sharply peaked, in the sense that its mean $b$ is much larger than its standard deviation, which is of order $(k^*)^{-1/2}$.  In this case, the distribution of $m_k$ also remains sharply peaked for the later part of the trajectory.  For $k>k^*$, the system follows the original dynamics and the mean of $m_k$ can be obtained from (\ref{equ:eom-m}) by solving
$
\mathrm{d}m/\mathrm{d}k = (\tanh Jm - m)/k
$, as in~\cite{kggw18}.
It is natural to change variables to $u=\log k$ so that 
\beq
\frac{\mathrm{d}m}{\mathrm{d}u} = \tanh Jm - m \; .
\label{equ:dmdu}
\eeq  
This means that for $k>k^*$, trajectories will flow away from the unstable fixed point at $m=0$ and towards the stable point at $m=\mS$, as in Fig.~\ref{fig:rateFn}(a).   Moreover, this evolution is very slow: the natural time variable is not the number of steps $k$ but the rescaled ``time'' $u=\log k$.  
(Physically, the slow variation with $k$ occurs because there are already many particles in the cluster, so making a significant change in its magnetisation requires the addition of many particles.)

In order to establish a bound on ${\cal I}_K(m,\eps)$, we treat $m$ as a target value for $m_K$: suitable controlled processes should hit this target with high probability.  
We choose $b$ and $k^*$ to achieve this, as follows.   

We solve 
(numerically) the differential equation (\ref{equ:dmdu}), going backwards in time.  This yields a path which ends at $m_K=m$ and can be propagated back to any earlier time $k$, for example by Euler's method.  The magnetisation on this path is $\tilde{b}(k)$ with $k\leq K$ and $\tilde{b}(K)=m$.    Both $k$ and $\tilde b(k)$ decrease as we solve the equation backwards in time.  As $k\to0$ then $\tilde b\to0$. (Note that $m=0$ is an unstable fixed point of the forward equation, which corresponds to a stable fixed point of the backward equation.)
We stop the solution at the point $(k,\tilde b)=(k^*,b)$ where 
 \beq
b^2 k^* = a\sqrt{K}
\label{equ:sqrt}
\eeq
where $a$ is a numerical parameter, of order unity (we take $a=2$, results depend weakly on this choice).   
  
 
%

These $b,k^*$ are the parameters that we use for the controlled dynamics.
As long as $K$ is reasonably large, the algorithm gives $b\ll 1$ and $k^*\gg 1$.
Then the action for this controlled process can be estimated from (\ref{equ:A-small-b}), with $K$ replaced by $k^*$.  This yields
\beq
\langle {\cal A} \rangle_{\rm con} \approx k^* \frac{b^2 (J-1)^2}{2} \; .
\label{equ:A-sqrtK}
\eeq
Since $(b,k^*)$ solve (\ref{equ:sqrt}), this means that $\langle {\cal A} \rangle_{\rm con}$ is of order $\sqrt{K}$.
Combining this result with (\ref{equ:def-H}) and assuming that the controlled system hits the target with probability 1, one obtains
\beq
{\cal H}_K(m,\eps) \simeq \frac{a(J-1)^2}{2\sqrt{K}} \; .
\label{equ:H-sqrtK}
\eeq
Hence, the bound ${\cal H}_K$ tends to zero as $K\to\infty$.  This result applies for large $K$ and is independent of the target chosen for $m_K$ (always assuming that this target is between $\eps$ and $\mS-\eps$).  Note however, that the controlled process depends on $K$, in that the parameters $b,k^*$ are chosen separately for each value of $K$.  The assumption that the controlled system hits the target with probability 1 is valid as long as the magnetisation distribution at $k^*$ is sharply-peaked in the sense that its mean is much larger than its standard deviation.  This requires $b^2k^*\gg1$ which is true by (\ref{equ:sqrt}) as long as $K$ is large.

Numerical results based on this construction are shown in Fig.~\ref{fig:rateFn}.  In particular, Fig.~\ref{fig:rateFn}(a) shows that the parameters $(b,k^*)$ obtained by this method are such that the controlled system hits the target $m$ with high probability.  Also, Fig.~\ref{fig:rateFn}(b) confirms that on increasing $K$,  the bound ${\cal H}_K$ is quantitatively consistent with (\ref{equ:H-sqrtK}). This establishes that $\lim_{K\to\infty} {\cal H}_K(m,\eps)=0$, and hence from (\ref{equ:IK-bound}) one has $\lim_{K\to\infty} {\cal I}_K(m,\eps)=0$, for this value of $m$.  Fig.~\ref{fig:rateFn}(c) shows that the same behaviour occurs for several values of $m$ with $|m|<\mS-\eps$.  This is expected since the theoretical argument above is independent of the target for $m_K$.  Hence $\lim_{K\to\infty} {\cal I}_K(m,\eps)=0$ throughout this range, which establishes that the rate function  (\ref{equ:ldp-mk}) obeys
\beq
I(m) = 0, \qquad |m| < \mS \; .
\eeq
This was the result anticipated in Fig.~\ref{fig:rate-analytic}.  
A similar result has been proven by a rigorous analysis of a general class of Polya urns, see Corollary 5 of~\cite{Franchini2017}.
We emphasise here that while the numerical results in Fig.~\ref{fig:rateFn} are a useful confirmation of our theoretical calculations, the bound in (\ref{equ:H-sqrtK}) is an analytical result.  

\subsection{Scaling of the probability that $|m_K|<\mS$}
\label{sec:beyond-ldp}

As $K\to\infty$, we have shown that ${\cal I}_K(m,\eps)\to0$ throughout the regime $|m|\leq \mS-\eps$.   The shows that $\mathrm{Prob}\!\left( |m_K-m| < \eps \right)$ does not decay exponentially with $K$.  Nevertheless, we expect that this probability should vanish as $K\to\infty$, so the natural question is: how small is it?  

To address this question we define an estimate of the probability density for $m_K$ as
\beq
\rho_K(m,\eps) = \frac{1}{2\eps} \mathrm{Prob}\!\left( |m_K-m| \leq \eps \right)
\label{equ:rho}
\eeq
(In this Section, we emphasise that all large-$K$ limits are to be taken at fixed $\eps>0$, note also that we are assuming $J>1$.)

\begin{figure}
\includegraphics[width=8.6cm]{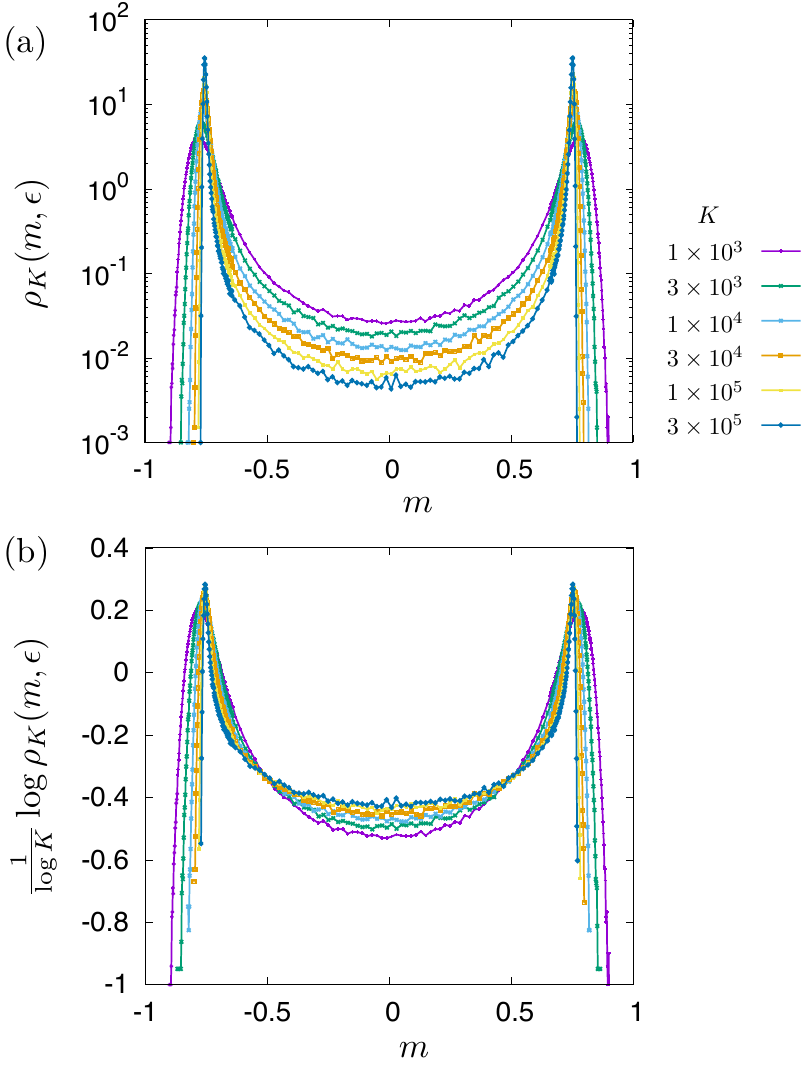}
\caption{(a) Estimated probability density for $m_K$ obtained by direct sampling of the irreversible growth model at $J=1.3$.  Note that $K$ varies over more than two decades.  The binning parameter is $\eps=10^{-3}$ for values of $m$ near the peaks and $\eps=10^{-2}$ for intermediate values.  (Results depend weakly on $\eps$, which is chosen to reduce statistical uncertainties while maintaining adequate resolution.)   (b) The logarithms of the same probability densities, scaled by $\log K$.  The prediction (\ref{equ:J-prob}) is that for every $|m|<\mS-\eps$ one should observe convergence of this quantity to some negative (non-zero) limit, as $K\to\infty$.  The data are consistent with this prediction.}
\label{fig:histo}
\end{figure}


Fig.~\ref{fig:histo}(a) shows the distribution of $m_K$, obtained by direct sampling of trajectories of the system, for the representative parameter value $J=1.3$.  Two features are notable.  First, the probability that $m_K\approx 0$ is small and decreases with $K$, but the decay is much slower than exponential in $K$, consistent with the arguments of section~\ref{sec:results-ldp-irrev}.   Second, there is no evidence for a local maximum in the probability at the unstable fixed point $m_K=0$.   We have verified that the behaviour for $|m|<\mS$ is similar for larger $J$, so this is a representative state point.  However, the behaviour close to the peaks of $\rho_K$ has a more complex dependence on $J$, see Sec.~\ref{sec:harris-regime} below.

%
%

To estimate $\rho_K$ for $|m|<\mS$, it is possible to repeat the argument of  Sec.~\ref{sec:results-ldp-irrev}, replacing (\ref{equ:sqrt}) with $b^2 k^* = aK^{\alpha}$ for any $\alpha\in(0,1)$.  This can be used to show that for any $\beta>0$ one has
\beq
\lim_{K\to\infty} K^{-\beta} \log \mathrm{Prob}\!\left( |m_K-m| < \eps \right) = 0 \;,
\eeq 
for $|m|\leq(\mS-\eps)$ as usual.   In other words, the probability of a non-typical value of $m_K$ decays slower than $\exp(-cK^\beta)$, for any $c,\beta>0$.  Based on this observation, we propose that  the probability decays as a power law in $K$.  In that case
\beq
{\cal J}_K(m,\eps) = -\frac{ \log  \rho_K(m,\eps) }{ \log K}
\label{equ:J-prob}
\eeq
should have a positive (non-zero) limit as $K\to\infty$.
Fig.~\ref{fig:histo}(b) shows results that are consistent with (\ref{equ:J-prob}).  We now present theoretical arguments that further support this conjecture, including bounds based on (\ref{equ:IK-bound}).

\begin{figure*}
\includegraphics[width=17.8cm]{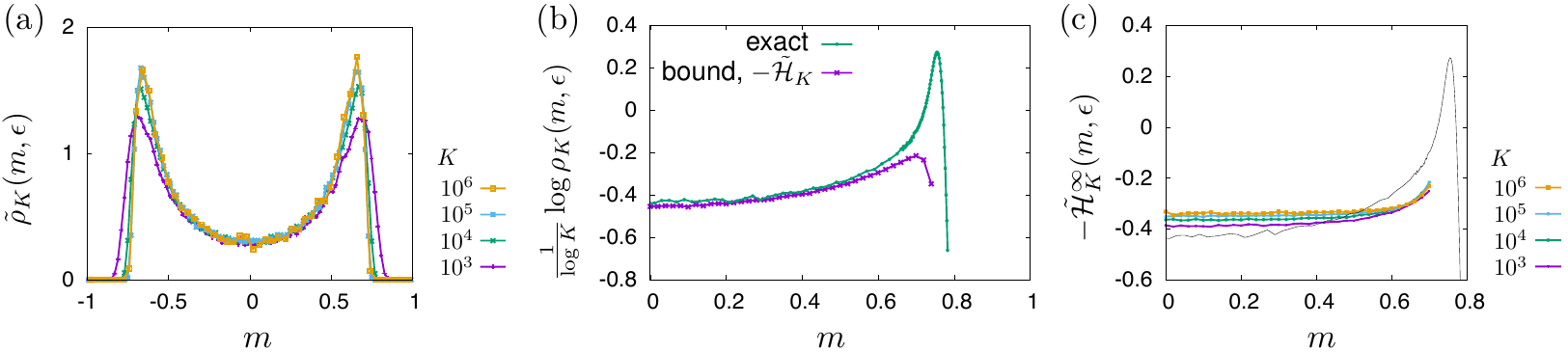}
\caption{Irreversible model with $J=1.3$.  (a) Probability distribution of $m_K$ under the controlled dynamics, with $k^*$ chosen as in (\ref{equ:kstar-good}).  The form of the distribution depends weakly on $K$ and the probability density is of order unity across a wide range of $m$. (b) The bound (\ref{equ:JK-bound}) for $K=10^5$, compared with the (numerically) exact distribution $\rho_K$ for the original model, obtained by direct sampling.  (c) The negative of the asymptotic bound $\tilde{\cal H}_K^\infty$ that appears in (\ref{equ:asymp-bound}),  for various $K$.  As an illustrative comparison, this is compared with the (numerically) exact distribution at $K=10^5$ from panel (b), shown as a thin black line. We emphasise that the asymptotic bound is not a bound on this finite-$K$ distribution}
\label{fig:bound-good}
\end{figure*}

\subsubsection{Accurate bounds for large $K$}

Consider the same controlled dynamics as in Sec.~\ref{sec:results-ldp-irrev}, but with $b=0$.  The remaining parameter is $k^*$: this means that the extensive magnetisation $M_k$ in the controlled process is a simple (unbiased) random walk for $k<k^*$. We use the original growth dynamics (\ref{equ:Mk+1}) for $k>k^*$.   In this case the distribution of $m_{k^*}$ has mean zero and its standard deviation is $(k^*)^{-1/2}$, by (\ref{equ:diff}).  We will take $k^*\gg 1$  so the distribution of $m_{k^*}$ is sharply-peaked in the sense that its variance is small compared to unity.  However, in contrast to Sec.~\ref{sec:results-ldp-irrev}, this distribution does not remain sharply-peaked under the time evolution (see below).

To fix a suitable value for $k^*$, it is useful to consider the deterministic evolution of the mean of $m_k$ for $k>k^*$, assuming that $m_{k^*}$ has a typical value of order $(k^*)^{-1/2}$.   Since $m$ is small, one may linearise (\ref{equ:dmdu}), leading to $(\mathrm{d}m/\mathrm{d}u) = (J-1)m$ and hence $m_k = m_{k^*} (k/k^*)^{J-1}$.  We choose $k^*$ in such a way that this deterministic equation gives $m_K=O(1)$.   This leads to
\beq
k^* = \gamma K^{ \frac{J-1}{J-(1/2)} }
\label{equ:kstar-good}
\eeq
with $\gamma$ a constant of order unity.  We take $\gamma=1$.  

By analogy with (\ref{equ:rho}), we define an estimate of the probability density for $m_K$ under the controlled dynamics as 
\beq
\tilde{\rho}_K(m,\eps) = \frac{1}{2\eps} \mathrm{Prob}_{\rm con}\!\left( |m_K-m| \leq \eps \right)
\label{equ:rho-tilde}
\eeq
where $\mathrm{Prob}_{\rm con}$ indicates a probability under the controlled dynamics.
Numerical results for $\tilde\rho$ are shown in Fig.~\ref{fig:bound-good}(a), for several values of $K$, always with $k^*$ chosen according to (\ref{equ:kstar-good}).  One sees that the distributions of $m_K$ are not sharply peaked. Instead $\tilde\rho$ is of order unity everywhere between $\pm\mS$.  Moreover, this distribution depends very weakly on $K$ (which varies over three decades).  This is due to the choice proposed in (\ref{equ:kstar-good}): 
the value of $\gamma$ is not important but the correct power-law exponent is essential.  (Different values of $\gamma$ lead to different distributions, but they are all similarly broad.)

Repeating the argument of Sec.~\ref{sec:control}, one obtains
\beq
{\cal J}_K(m,\eps) \leq \tilde{\cal H}_K(m,\eps)
\label{equ:JK-bound}
\eeq
with
\begin{multline}
\tilde{\cal H}_K(\rho,\eps) = \frac{-1}{\log K} \log \tilde{\rho}_K(m,\eps) 
\\
+ 
 \frac{1}{\log K}  \frac{ \langle \chi_\eps(m_K-m) {\cal A}[\bm{m}] \rangle_{\rm con} }{\langle  \chi_\eps(m_K-m) \rangle_{\rm con}} \; .
 \label{equ:tildeH}
\end{multline}
This bound is shown in Fig.~\ref{fig:bound-good}(b), for the representative case $K=10^5$.  It shows almost quantitative agreement over a range of $m$ that includes $m=0$.  However, the agreement breaks down as $m$ gets close to $\mS$.  (Better bounds for larger $m$ might be obtained by using smaller $c$ in (\ref{equ:kstar-good}), but we have not explored this in detail.  See also Sec.~\ref{sec:harris-regime}.)  

\subsubsection{Asymptotic behaviour as $(\log K)\to\infty$}

The numerical bound in Fig.~\ref{fig:bound-good}(b) is valid at $K=10^5$ and gives a reasonable estimate of the probability density, but the regime of primary interest is $K\to\infty$.  Based on Fig.~\ref{fig:bound-good}(a), we expect that $ \log \tilde{\rho}_K(m,\eps)$ remains of order unity as $K\to\infty$.  In this case, its contribution to $\tilde{\cal H}_K$ will decay proportional to $1/(\log K)$ and the term involving ${\cal A}$ will dominate  (\ref{equ:tildeH}) at large $K$.  
It is useful to estimate this term directly, in order to predict the behaviour of $\tilde{\cal H}_K$ as $K\to\infty$.  That is, assuming that $\log \tilde\rho$ remains finite as $K\to\infty$, one has from (\ref{equ:JK-bound},\ref{equ:tildeH}) that
\beq
\lim_{K\to\infty} {\cal J}_K(m,\eps) \leq \lim_{K\to\infty} 
\tilde{\cal H}_K^\infty(m,\eps)
\label{equ:asymp-bound}
\eeq
with
\beq
\tilde{\cal H}_K^\infty(m,\eps)
=
\frac{1}{\log K} 
\frac{ \langle \chi_\eps(m_K-m) {\cal A}[\bm{m}] \rangle_{\rm con} }{\langle  \chi_\eps(m_K-m) \rangle_{\rm con}}  \; .
\label{equ:asymp-bound-val}
\eeq
We refer to this as an asymptotic bound since it does not bound ${\cal J}_K$ at any finite $K$, but only as $K\to\infty$: see (\ref{equ:asymp-bound}).  For finite $K$, the difference between the exact bound and the asymptotic bound scales as $(1/\log K)$, which decays very slowly with $K$.  This means that the asymptotic behaviour is not easily accessible in numerics.
However, the asymptotic bound can be evaluated numerically for finite $K$ and provides an estimate of the right hand side of (\ref{equ:asymp-bound}).

Results are shown in Fig.~\ref{fig:bound-good}(c). The estimates for the asymptotic bound are compared with the actual distribution for $K=10^5$.  
%
One sees that the asymptotic bound in Fig.~\ref{fig:bound-good}(c) depends weakly on $m$, over a fairly wide range that includes $m=0$.  This bound is evaluated numerically as an average of ${\cal A}$, which is restricted to trajectories whose final magnetisation $m_K$ is within the relevant bin of a suitably constructed histogram.  
We do not have a theoretical estimate of this conditional average.  However, the unconditioned average of the action can be obtained
from (\ref{equ:def-A}) as
\beq
\langle {\cal A}(\bm{m}) \rangle_{\rm con} \simeq \sum_{k=1}^{k^*} \left\langle \frac{1}{2}  \log \frac{(1+\ee^{-2Jm_k})(1+\ee^{2Jm_k})}{4} \right\rangle_{\rm con}
\label{equ:A-bzero}
\eeq
For $1\ll k \leq k^*$ then $m_k$ is small under the controlled dynamics and its variance is $1/k$.  Hence, we expand (\ref{equ:A-bzero}) to second order in $m_k$, which yields $\langle {\cal A}(\bm{m}) \rangle_{\rm con} \approx \sum_k J^2/(2k)$.   Approximating the sum by an integral leads to
\beq
\langle {\cal A}(\bm{m}) \rangle_{\rm con} \simeq \frac{J^2}{2} \log k^* + A_0 \; .
\label{equ:A-log}
\eeq
where $A_0$ is a constant of order unity which depends on the behaviour of trajectories when $k$ is of order unity. (The behaviour of the system for $k=O(1)$  is not captured by our various approximations so we are not able to estimate this constant analytically.)


If we now assume that the the conditional average of the action in (\ref{equ:def-H}) is 
is the same as the corresponding unconditioned average then (\ref{equ:kstar-good},\ref{equ:tildeH},\ref{equ:A-log}) together yield 
\beq
\tilde{\cal H}^\infty_K(m,\eps) \simeq \frac{J^2 (J-1)}{2J-1} \; .
\label{equ:Hinf-J}
\eeq
which is independent of $m$.  Using the unconditioned average as an estimate of the conditioned one is an uncontrolled approximation,
but (\ref{equ:Hinf-J}) is consistent with the weak dependence of $\tilde{H}^\infty_K$ on $m$ in Fig.~\ref{fig:bound-good}(c).  For that case, (\ref{equ:Hinf-J})
evaluates to 0.32, consistent with the data for small and moderate $m$ (given that we are still far from the limit where $\log K$ is large).
%

Based on this analysis, we summarise our conclusions for the probability distribution of $m_K$, at large $K$.
We have argued that for $|m|<(\mS-\eps)$ then ${\cal J}_K(m,\eps)$ has a finite limit as $K\to\infty$, which means that $\rho_K(m,\eps)$ decays as a power law.
Based on the asymptotic bound in Fig.~\ref{fig:bound-good}c, we offer two possibilities for the detailed behaviour of ${\cal J}_K$.  The first is that $\lim_{K\to\infty} {\cal J}_K(m,\eps)$ is independent of $m$ within some range including zero, and that it takes a value $\alpha$ within that range. From (\ref{equ:Hinf-J}), we expect $\alpha \approx J^2 (J-1)/(2J-1)$ in this case. The result would be that
\beq
\rho_K(m,\eps) \simeq K^{-\alpha} f(m,\eps)
\label{equ:no-ldp}
\eeq
within the relevant range of $m$, with $f(m,\eps)$ of order unity.  The extreme version of this scenario would be that the relevant range is $|m|<\mS-\eps$: this is not apparent from the finite-$K$ data presented here but we are still far from the limit where $\log K\to\infty$.  In any case this limit is not expected to commute with a limit where $m\to(\mS-\eps)$ so one may expect significant deviations from the asymptotic (large-$K$) behaviour in data obtained at finite $K$.
The second scenario is that $\lim_{K\to\infty} {\cal J}_K(m,\eps)=\alpha(m,\eps)$ where $\alpha$ is now a function which takes values of order unity. In this case
\beq
\rho_K(m,\eps) \simeq K^{-\alpha(m,\eps)}  \; .
\label{equ:ldp-logt}
\eeq
This might be interpreted as a large deviation principle with speed $(\log K)$, because the distribution of $m_K$ scales as $p(m_K) \sim \ee^{-(\log K) \beta(m_K) }$, where $\beta$ would be interpreted as a rate function [with $\beta(m)=\lim_{\eps\to0}\alpha(m,\eps)$].  However, establishing such an LDP would require a detailed mathematical analysis that is beyond the scope of this work.

 Based on the available numerical data, we are not able to settle which (if either) of (\ref{equ:no-ldp},\ref{equ:ldp-logt}) is valid, because all results are limited by the fact that the numerical parameter $\log K$ governs the convergence to the large-$K$ regime, and this number is never very large.  
 This might be addressed by noting that for large $k$, the change in $m_k$ on any single step is very small, so one might promote both $k$ and $m_k$ to continuous variables, and describe the discrete-time stochastic process (\ref{equ:Mk+1}) by a stochastic differential equation.  Such a construction would be similar to the temporal addivity principle of~\cite{Harris2009}, but requires a detailed analysis that is beyond the scope of this work.

 \subsection{Scaling of the probability distribution of $m_K$ for $|m_K|\gtrsim\mS$}
 \label{sec:harris-regime}

We have shown how control forces that depend on $k$ can be used to derive bounds on the rate function,  and that these bounds differ qualitatively from the (loose) upper bounds that are obtained using control forces that are independent of $k$~\cite{kggw18}.  All results so far are relevant for the probability distribution of $m_K$ when $|m_K|<\mS$, which is the main focus of this paper.

This Section discusses the behaviour for $|m_K|\gtrsim\mS$, based on the theory of Harris~\cite{Harris2015}.
We show how $k$-dependent control forces can yield improved bounds on probability distributions in that case too.  
In particular, we analyse the behaviour of probability distributions such as the one in Fig.~\ref{fig:histo}, for values of $m$ close to the peaks.

Before presenting the calculation, we first summarise the behaviour that would usually be expected~\cite{Touchette2009} based on the LDP (\ref{equ:ldp-mk}).  One would expect that a central limit theorem (CLT) holds, so that the sharp peaks of $\rho_K(m,\eps)$ have width (standard deviation) $\sigma K^{-1/2}$ with $\sigma=I''(\mS)^{-1/2}$.  However, for $J>1$, one sees from Fig.~\ref{fig:rate-analytic} that $I''(m)$ is discontinuous at $\mS$.  In this situation, one expects in general that $\sigma=I''(\mS^+)^{-1/2}$ where the notation indicates that one should evaluate the derivative as $m\to\mS^+$. 
The following analysis shows how the behaviour of the growth model differs from this simple picture.
 
\subsubsection{Control forces with arbitrary time-dependence}

We apply the theory of~\cite{Harris2015,Harris2009} by connecting it to the controlled process introduced in Equ.~(\ref{equ:Mk+1-control}).  
This requires that we consider $k$-dependent control forces where the parameter $b$ has an arbitrary dependence on $k$, in contrast to the simple choices considered so far.  This leads to $\langle M_k \rangle_{\rm con} = \sum_{x=1}^k b_{x}$ where the sum runs over steps. Promoting $k$ to a continuous variable we define 
\beq
m(k) = \langle m_k \rangle_{\rm con} = \frac{1}{k} \int_0^k b(x) \mathrm{d}x
\eeq
and one sees that $m'(k) = [b(k) - m(k)]/k$ where the prime denotes a derivative with respect to $k$.  Hence, for any (smooth) path $m(k)$, we can define a
a controlled process that follows this path by taking
\beq
b(k) = m(k) + km'(k) \; .
\label{equ:b-mprime}
\eeq
Following (\ref{equ:A-homog},\ref{equ:HK-homog}), the action for such a time-dependent path can be estimated as
\begin{multline}
\langle {\cal A}\rangle_{\rm con} \approx \int_0^K \Big\{ \frac12 \log [1-b(k)^2] + \log \cosh(Jm(k))
\\ + \frac{b(k)}{2}  \log \frac{1+b(k)}{1-b(k)} - Jm(k)b(k) \Big\} \mathrm{d}k \; ,
\label{equ:A-gen-kdep}
\end{multline}
%
which assumes that the distribution of $m_k$ is sharply peaked and is valid up to an additive constant of order $\log K$.
To obtain the best available bound ${\cal H}_K(m,\eps)$ via (\ref{equ:def-H}), this action should now be minimised over the path $m(k)$, subject to (\ref{equ:b-mprime}) and $m(K)=m$.

Here, we take a simpler route.  Let $m^*$ be a stable fixed point of the deterministic dynamics, so $m^*=0$ for $J<1$ and $m^*=\pm\mS$ for $J>1$. We restrict our analysis to paths for which $m(k)-m^*$ and $b(k)-m^*$ are both small compared to unity.   This allows us to apply the results of~\cite{Harris2015}.  
Under these assumptions the action (\ref{equ:A-gen-kdep}) may be expanded as
\beq
\langle {\cal A}\rangle_{\rm con} \approx \frac{1}{2\chi} \int_0^K \left[ km'(k) + (m(k)-m^*)\Lambda \right]^2 \mathrm{d}k
\label{equ:A-quad}
\eeq
where we used (\ref{equ:b-mprime}); and $\Lambda = 1-J[1-(m^*)^2]$ and $\chi=1-(m^*)^2$ are numerical parameters that correspond to $(1-A^*)$ and $D^*$ in Equ.~(7) of Ref.~\cite{Harris2015}.
We neglected contributions at ${\cal O}(m-m^*,b-m^*)^3$ in the integrand.
For forces that are independent of $k$ the action is $ \langle {\cal A}\rangle_{\rm con} \approx {\cal A}_{\rm ind} $ with 
\beq
{\cal A}_{\rm ind} = \frac{\Lambda^2}{2\chi} ( m - m^* )^2 K \; .
\label{equ:A-ind}
\eeq
Using this result with (\ref{equ:def-H},\ref{equ:IK-bound}) yields a bound on ${\cal I}_K$ which was also derived in KGGW~\cite{kggw18}.

The next step is to minimise the action  $ \langle {\cal A}\rangle_{\rm con} $ over paths $m(k)$ that satisfy $m(K)=m$.  The resulting action yields a bound on ${\cal I}_K(m,\eps)$.  
We refer to Ref.~\cite{Harris2015} for all details of the minimisation procedure.  
The important physical conclusion is that the solutions do not have constant forces, so one finds  in general that 
$\langle {\cal A}\rangle_{\rm con} < {\cal A}_{\rm ind}$.

The reduction in the action  
depends strongly on the parameter $\Lambda$.  One always has $0<\Lambda<1$, and the results depend qualitatively on whether $\Lambda$ is bigger or smaller than $\frac12$~\cite{Harris2015}.  Small values of $\Lambda$ correspond to systems with large fluctuations.  

\subsubsection{Results for the growth model}

In the irreversible growth model, we find that $\Lambda>\frac12$ if the parameter $J$ is far from its critical value $J_c=1$.  
Specifically, $\Lambda>\frac12$ if either $J<0.5$ or $J>1.37$.
In these cases, the theory~\cite{Harris2015} gives a minimal action
\beq
\langle {\cal A}\rangle_{\rm con} \approx \frac{2\Lambda-1}{2\chi} ( m - m^* )^2 K \; .
\label{equ:Acon-Lambda-big}
\eeq
Comparing with (\ref{equ:A-ind}), one sees that introducing $k$-dependent control forces reduces the action by a factor of $(2\Lambda-1)/\Lambda^2$.

However, it is important to consider the optimal paths that are predicted by this theory.
For convenience we restrict to the case where $m,m^*>0$.  The optimal paths take $m(k)=m^*$ for $k<k_*$ where $k_*\ll K$ is a parameter.  
For $k>k_*$ then $m(k)$ has an excursion away from $m^*$, it grows to a value of order $(K/k_*)^{1-\Lambda}(m-m^*)$  before converging towards the target as $m(k) -m^* \approx (K/k)^{1-\Lambda} (m-m^*)$.
Note however that we required $m(k)-m^*\ll 1$ when deriving (\ref{equ:A-quad}): hence one must have $(m-m^*) = {\cal O}(k_*/K)^{1-\Lambda}$.  Since the derivation of~(\ref{equ:Acon-Lambda-big}) also requires that $k_*\ll K$, one sees that $m-m^*$ must be small.

\begin{figure}
\includegraphics[width=70mm]{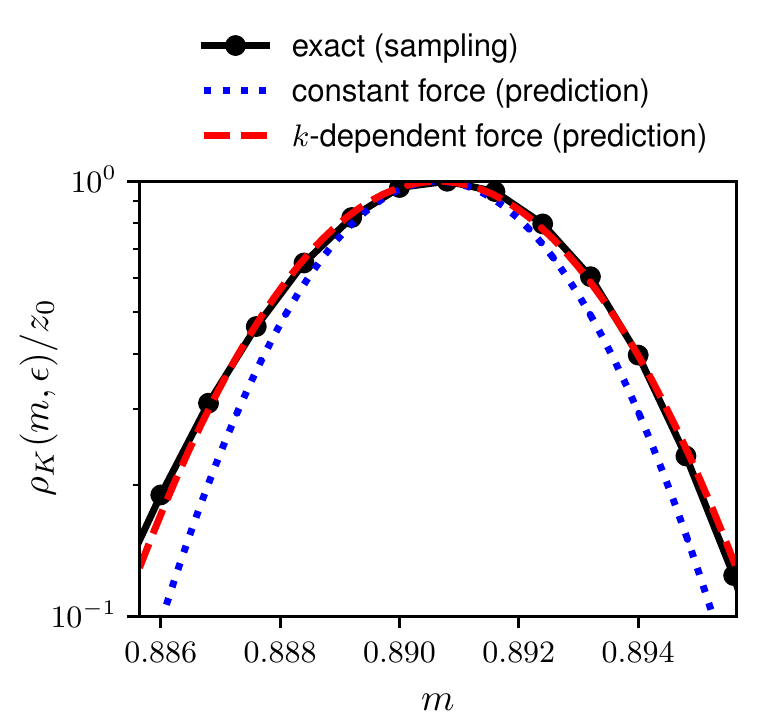}
\caption{Probability distribution $\rho_K(m,\eps)$ for $J=1.6$ with $K=10^5$ and $\eps=8\times10^{-4}$, showing the behaviour close to the peak at $m^*=\mS\approx0.8906$.  In this case $\Lambda\approx0.67$.  The probability density is shown on a logarithmic scale; it has been divided by $z_0=\rho_K(m^*,\eps)$ so that the maximal value is unity.  Numerical data were obtained by direct sampling and are compared with the predictions obtained using constant control forces [Equ.~(\ref{equ:A-ind})] and $k$-dependent control forces [Equ.~(\ref{equ:clt-lambda-big})].  There are no fitting parameters.  The predictions are accurate if their respective controlled processes capture the typical mechanism for fluctuations of $m_K$; they also require that $(m-m^*)$ is small.  The controlled process with constant forces does not fully capture the fluctuation mechanism  so it underestimates the relevant probability.  The process with $k$-dependent forces captures the mechanism and gives accurate predictions close to the peak, but it does not capture the (mild) skewness of the exact distribution, because the Taylor expansion in (\ref{equ:A-quad}) is truncated at second order.}
\label{fig:J-16-peak}
\end{figure}

For this reason, we focus on values of $m$ that are very close to the peaks of $\rho_K(m,\eps)$.  We take $m-m^*=vK^{-x}$ for parameters $v,x>0$, so the difference between $m$ and $m^*$ vanishes as $K\to\infty$.  In addition we assume that $x\geq(1-\Lambda)$: this includes the possibility that $x=\frac12$ because $\Lambda>\frac12$.  In this case our various assumptions are all satisfied and (\ref{equ:Acon-Lambda-big}) yields
\beq
\langle {\cal A}\rangle_{\rm con} \approx \frac{2\Lambda-1}{2\chi} v^2 K^{1-2x} \; ,
\label{equ:Acon-v-lambda-big}
\eeq
independent of $k_*$.  Since all the assumptions of~\cite{Harris2015} are satisfied, we expect that this action is optimal in the sense that substituting in (\ref{equ:def-H}) should yield a bound ${\cal H}_K$ that achieves equality in (\ref{equ:IK-bound}), as $K\to\infty$.  The case $x=\frac12$ is instructive because the resulting action is of order unity as $K\to\infty$: this indicates that fluctuations on this scale are typical and that the peaks of $\rho_K(m,\eps)$ can be approximated by Gaussian probability densities.  The variances of these Gaussian peaks are predicted to be
\beq
{\rm Var}(m_K) \simeq \frac{\chi}{(2\Lambda-1)K}   \; .
\label{equ:clt-lambda-big}
\eeq
(Note, in contrast to previous results that described large deviations, this result is relevant for typical fluctuations.)

As an illustrative example of the behaviour near the peak of $\rho_K$, Fig.~\ref{fig:J-16-peak} shows results for $J=1.6$, for which  $\Lambda\approx0.67$.  The data obtained by direct sampling are well-described by a Gaussian distribution with variance as in (\ref{equ:clt-lambda-big}), which differs significantly from the corresponding prediction that was obtained using control forces  that are independent of $k$. (This is the prediction obtained by using (\ref{equ:A-ind}) as a bound and assuming that this bound coincides with the rate function).  

Note that the dependence of (\ref{equ:clt-lambda-big}) on $K$ resembles a CLT.  However, it describes the behaviour of a single peak, while the distribution $\rho_K$ has two peaks; the true variance of $m_K$ is close to $\mS^2$ which is of order unity.  It is also possible to consider the distribution of $|m_K|$, which has a unimodal probability density function whose peak is well-described by a Gaussian with variance (\ref{equ:clt-lambda-big}), recall Fig.~\ref{fig:J-16-peak}.  In this case we still find numerically (for $J=1.6$) that the true variance of $|m_K|$ is much larger than (\ref{equ:clt-lambda-big}).  This effect can be explained using the results of Sec.~\ref{sec:beyond-ldp}: the distribution of $|m_K|$ has a heavy tail that extends all the way to $|m_K|=0$, and this has a significant contribution to the variance (recall Fig.~\ref{fig:histo}).  
We conclude that (\ref{equ:clt-lambda-big}) predicts the width of the peak of $\rho_K(m,\eps)$ but it is not sufficient to establish a CLT, neither for $m_K$ nor for $|m_K|$.  On the other hand, for $J<0.5$ there is no heavy tail and we do expect a CLT for $m_K$.  The results of Sec.~\ref{sec:beyond-ldp} also indicate that a CLT for $|m_K|$ might be expected for larger values of $J$, but this is beyond the scope of this work.

%

The result (\ref{equ:clt-lambda-big}) raises the question of what happens as $\Lambda\to\frac12 $, where this variance seems to diverge.
To understand this, we consider the case $\Lambda<\frac12$,
for which the physics changes qualitatively.  Assuming as before that $K\gg k_*$, the analogue of (\ref{equ:Acon-Lambda-big}) is~\cite{Harris2015}  
\beq
\langle {\cal A}\rangle_{\rm con}  =\frac{1-2\Lambda}{2\chi}(m-m^*)^2  \cdot k_*^{1-2\Lambda}  \cdot K^{2\Lambda} \; .
\label{equ:Acon-Lambda-small}
\eeq
%
In the optimal path that gives this result~\cite{Harris2015}, the maximal value of  $m(k)$ is of order $(K/k_*)^{\Lambda}(m-m^*)$. Following the same procedure as before, this suggests that we take $m-m^*=vK^{-x}$ with the restriction $x\geq\Lambda$ (which again includes $x=\frac12$ because we are now considering $\Lambda<\frac12$).  The result is that
\beq
\langle {\cal A}\rangle_{\rm con} = \frac{1}{2\chi} (1-2\Lambda) v^2 \cdot K^{2(\Lambda-x)} \cdot k_*^{1-2\Lambda} \; .
\eeq
Note that $k_*$ is still a free parameter which we assume to be of order unity.  For the case $x=\frac12$ one sees that $\langle {\cal A}\rangle_{\rm con}\to0$ for small $K$, in contrast to $\langle {\cal A}\rangle_{\rm con}={\cal O}(1)$ in (\ref{equ:Acon-v-lambda-big}).  In fact $\langle {\cal A}\rangle_{\rm con}\to0$ for all $x<\Lambda$, independent of $k_*$.  For $x=\Lambda$ we expect an action of order unity.  Obtaining its numerical value would require optimisation of the parameter $k_*$, but this depends on contributions to the action from steps with $k=O(1)$, which are not captured by our various approximations.  However, since the action is of order unity for fluctuations of this size, so we expect that $\rho_K(m,\eps)$ has peaks with width of order $K^{-\Lambda}$.  That is,
\beq
{\rm Var}(m_K) = {\cal O}(K^{-2\Lambda})   \; ,
\label{equ:clt-lambda-small}
\eeq
which corresponds to peaks that are much broader than (\ref{equ:clt-lambda-big}).  
This is consistent with (\ref{equ:clt-lambda-big}): the apparent divergence of that variance as $\Lambda\to\frac12$ signals a change in scaling behaviour from $K^{-1}$ to $K^{-2\Lambda}$.  This means that the variance always tends to zero on taking $K\to\infty$ at fixed $\Lambda$.
Equ.~(\ref{equ:clt-lambda-small}) should apply (for example) to the variances of the individual peaks in Fig.~\ref{fig:histo}, since $\Lambda\approx 0.43<\frac12$ in that case.  We also predict that it should hold in the one-phase regime, for $0.5<J<1$: this behaviour was demonstrated for the (so-called) artificial model of~\cite{Harris2015-peak-end}, whose behaviour is closely related to the model considered here.  A more detailed analysis of (\ref{equ:clt-lambda-small}) for the growth model would include numerical tests and determination of the prefactor in the scaling law, but this is beyond the scope of this paper.

In  addition to the results derived so far,
the arguments of~\cite{Harris2015} also suggest that the behaviour in (\ref{equ:clt-lambda-small}) should be associated with a new LDP whose speed is less than $K$.  
The calculations presented here are not sufficient to demonstrate this, because our analysis of optimal paths is restricted to very small $m-m^*$.  However, we summarise the physical picture that may be expected, if the arguments of~\cite{Harris2015} can be extended to situations where $m-m^*={\cal O}(1)$, for this model.
For $\Lambda>\frac12$ one expects the LDP (\ref{equ:ldp-mk}) and also (\ref{equ:clt-lambda-big}) to hold.
For $\Lambda<\frac12$ then one expects from (\ref{equ:Acon-Lambda-small}) that $I(m)=0$ for all $m$ and one should have a different LDP with speed $K^{2\Lambda}$:
\beq
p(m_K) \simeq \exp\left[ -K^{2\Lambda} I_\Lambda(m_K) \right] \; .
\eeq
In this case, (\ref{equ:clt-lambda-small}) arises naturally, and the numerical value of the variance is related to the second derivative of $I_\Lambda$.
For $J>1$ and $|m_K|<\mS$, we recall from Sec.~\ref{sec:beyond-ldp} that $p(m_K)$ always decays as a power law in $K$, which means that one still has $I_\Lambda(m_K)=0$ in this range.  Hence the rate function $I_\Lambda$ would be qualitatively similar to Fig.~\ref{fig:rate-analytic}. 

This is an appealing  physical picture, which is consistent with the general expectations of~\cite{Harris2015}.  However, we emphasise again that it would require significant further mathematical analysis to confirm it.  For example, our results here are also consistent with a scenario where (\ref{equ:ldp-mk}) holds for all values of $\Lambda$, with $I(m)>0$ for all $|m|>\mS$.  In this case (\ref{equ:clt-lambda-small}) would imply that $I''(0)=0$ for $0.5<J<1$ and $I''(\mS^+)=0$ for $1<J<1.37$.  For other values of $J$ then these second derivatives would be presumably be non-zero.

\section{Results : Reversible model}
\label{sec:rev-results}

In this section we analyse large deviations in the reversible model of growth, which was defined in Sec.~\ref{sec:rev-model}.  This demonstrates that the use of time-dependent control forces to obtain improved bounds can be generalised beyond the single system considered so far, leading to results similar to Fig.~\ref{fig:rate-analytic}.  We focus on a representative state point where the  deterministic dynamics has unstable fixed points at non-zero magnetisation. This is the case where the model has three steady states.  (For parameters where the model has one steady state or two steady states, the behaviour is very similar to that of the irreversible model, so we do not discuss it in detail.)

\subsection{Summary of behaviour and definition of controlled dynamics}

We summarise the behaviour of the model, as derived in~\cite{Klymko2017}.  We assume throughout that the parameter $c$ in (\ref{equ:Mk+1-rev}) is large enough that the steady state of the system involves a cluster that is grows with a constant rate.   (The alternative is that there is a steady state with a cluster of finite size, the behaviour is quite different in that case and we do not discuss it here.)  
For the reversible model, the analogue of (\ref{equ:eom-m}) is
\beq
\langle \Delta m_k \rangle_{m_k,N_k} 
= 
 \frac{(1- m_k ^2) \sinh (J m_k) - c  m_k }
 { N_k \left[ c + \cosh (J m_k) -  m_k \sinh (J m_k) \right] } \; .
\label{equ:eom-m-rev}
\eeq
where the average is conditioned on the values of $m_k$ and $N_k$.
As noted in~\cite{Klymko2017}, this means that steady state values for $m$ solve
\beq cm = (1-m^2) \sinh Jm \label{equ:cm-sinhM} \; . \eeq  
For $J>c$ there are three solutions to this equation, and the behaviour is similar to the irreversible model for $J>1$.  For $J<c$ there are two possibilities -- either a single solution, or five solutions.  The first case corresponds to a single steady state with $m=0$.   The latter case means that there are three possible steady states, which have magnetisations $0,\pm \mS$.  In this case there are also two unstable fixed points of (\ref{equ:eom-m-rev}) at $\pm m_{\rm U}$ (these are not steady states of the stochastic model, similar to the irreversible case).  The situation with three steady states is only possible for $\sqrt{6}<J<c$ (this condition is necessary but not sufficient).

As a suitable controlled dynamics, we take
\beq
(\Delta N_k,\Delta M_k) = 
\begin{cases} 
(+1,+1), &  \!\!\hbox{w/prob } (c'/z'_k)
\\
(+1,-1), &  \!\!\hbox{w/prob } (c'/z'_k)
\\
(-1,+1), &  \!\!\hbox{w/prob } (1+\lambda) (1-m_k) / z_k'
\\
(-1,-1), &  \!\!\hbox{w/prob } (1-\lambda) (1+m_k) / z_k'
\end{cases}
\label{equ:Mk+1-rev-control}
\eeq
with two parameters $\lambda,c'$, also $z_k' = 2(c'+1-\lambda m_k)$ for normalisation.  This controlled process is similar to that considered in~\cite{kggw18}, but not identical. In principle one may take different rates for the two cases with $\Delta N_k=+1$, which will lead to improved bounds in some cases.  However, (\ref{equ:Mk+1-rev-control}) is sufficient for our purpose, which is to show that $I(m)=0$ for $|m|\leq \mS$.

In the steady state of this model, the rate of cluster growth is denoted by $r$ and the magnetisation is denoted by $b$ (by analogy with (\ref{equ:diff})).  
 Considering the mean increments for $N_k$ and $M_k$, the rate of cluster growth can be verified to be $r=(c'-1+\lambda b)/(c'+1-\lambda b)$, which is assumed to be positive, as noted above. Also, $b$ is a solution of $\lambda(b^2-1) + bc' = 0$, specifically 
\beq 
b=\frac{1}{2\lambda} \left( \sqrt{c'^2+4\lambda^2} - c' \right). 
\label{equ:b-ss}
\eeq
If $\lambda=0$ (equal probabilities for unbinding of up and down spins) then $b=0$ (no magnetisation in steady state).
 
 Using that the cluster size and the magnetisation are both sharply peaked in the steady state, the analogue of (\ref{equ:A-homog}) may be expressed as 
\begin{multline}
\langle {\cal A}(\bm{m}) /K \rangle_{\rm con} \simeq 
\log (z/z'_b)   + (2c'/z'_b) \log (c'/c)
\\ +
 \frac{1-b\lambda}{z'_b} \log (1-\lambda^2) 
 \\ 
  + \frac{\lambda - b}{z'_b} \left[  \log \frac{1+\lambda}{1-\lambda}  - 2Jb \right]
\label{equ:A-homog-rev2}
\end{multline}
with $z'_b=2(c'+1-\lambda b)$.  We emphasise that this formula is valid only if $b,c',\lambda$ are related by (\ref{equ:b-ss}), so that $b$ is the magnetisation of the steady state of the controlled process.  

\begin{figure*}
\includegraphics[width=16cm]{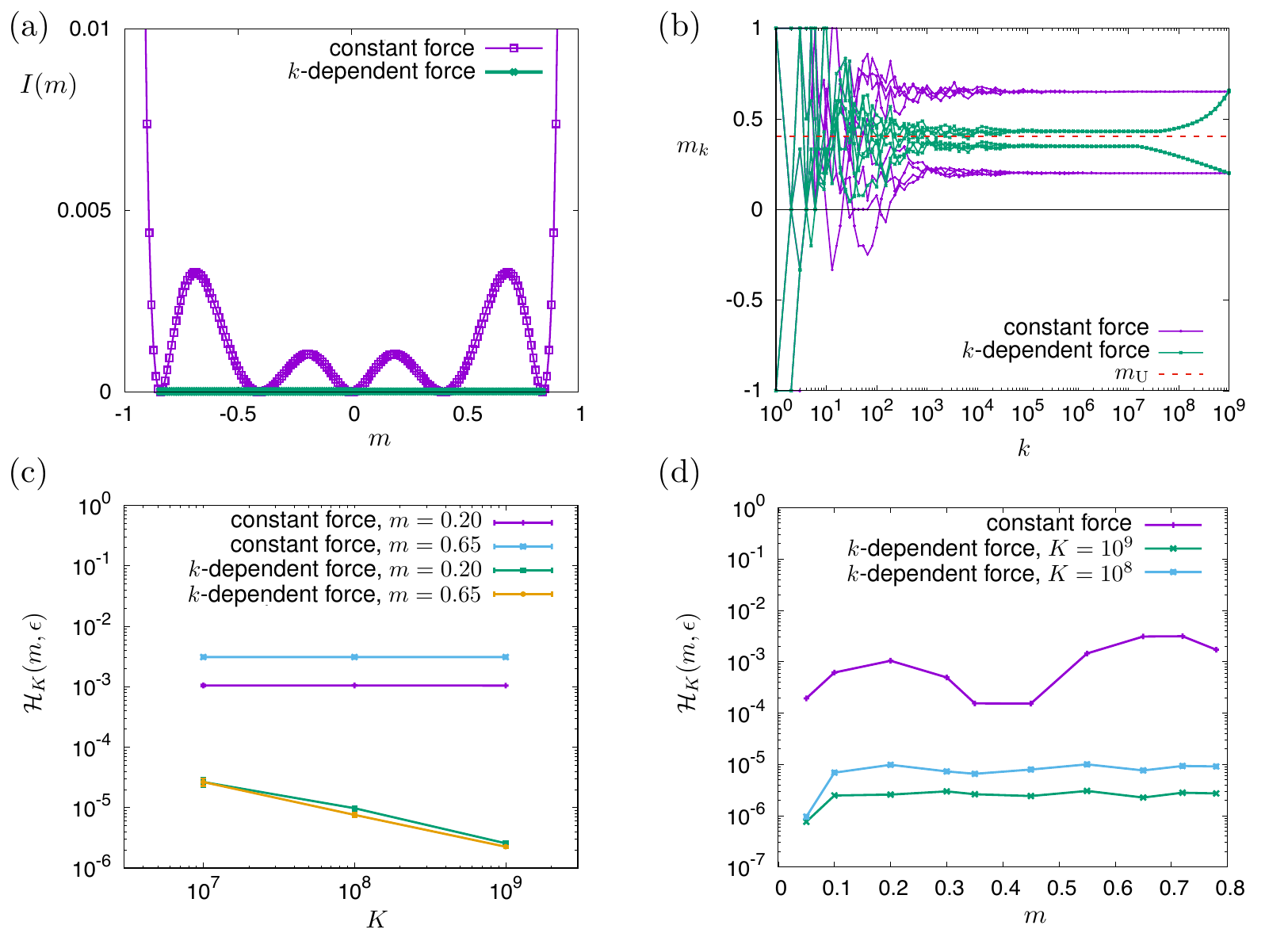} 
\caption{Reversible growth model with $(c,J)=(5.0,4.0)$, which is representative of the regime where the model has three steady states.  
(a)~Bounds on the rate function in (\ref{equ:ldp-rev}), obtained using control forces that are independent of $k$, and with $k$-dependent forces. Compare with Fig.~\ref{fig:rate-analytic}.
The data with $I(m)>0$ are obtained by numerical minimisation of (\ref{equ:A-homog-rev2}) but do not involve simulations of the growth process itself.
(b)~Trajectories of the controlled dynamics for $K=10^9$, which achieve either $m_K\approx0.2$ or $m_K\approx 0.65$. 
(c)~Results for ${\cal H}_K$ using these choices for the controlled dynamics.  Data are shown for increasing $K$, analogous to Fig.~\ref{fig:rateFn}(b).  For the time-dependent control force, this bound decays as $K^{-1/2}$, consistent with (\ref{equ:action-target}). (d)~The behaviour of ${\cal H}_K$ as a function of $m$, analogous to Fig.~\ref{fig:rateFn}(c).  The decrease of ${\cal H}_K$ with $K$ occurs for all $m$ within this range.
}
\label{fig:rateFnRev}
\end{figure*}

\subsection{Bounds on ${\cal I}_K(m,\eps)$}

The main result of this section is that ${\cal I}_K(m,\eps)\to0$ for $|m|<\mS$ and $K\to\infty$, and hence that the rate function $I(m)$ in (\ref{equ:ldp-rev}) is zero thoughout this range.  This is illustrated in Fig.~\ref{fig:rateFnRev}(a), which is similar to Fig.~\ref{fig:rate-analytic}.  As in the irreversible case, we prove this by obtaining bounds on ${\cal I}_K$.

As in Sec.~\ref{sec:control-kggw}, we first consider controlled processes with constant rates (independent of $k$).  This will recover results similar to KGGW.  We use (\ref{equ:A-homog-rev2}) with (\ref{equ:IK-bound}) to derive bounds on ${\cal I}_K(m,\eps)$, for $m=b$. 
 Using 
 \beq
 c'=\lambda(1-b^2)/b
 \label{equ:cprime}
 \eeq
  from above and also substituting for $z'_b$, the formula (\ref{equ:A-homog-rev2}) for the action may be expressed as a function of $(b,\lambda)$.  Minimising this function (numerically) over $\lambda$, one obtains a bound on ${\cal I}_K$, which is plotted with squares in Fig~\ref{fig:rateFnRev}(a).  
 Note that if $m=b$ is a solution of (\ref{equ:cm-sinhM}) then one may take $\lambda = \tanh Jb$ and $c'=c/(\cosh Jb)$, and the action (\ref{equ:A-homog-rev2}) evaluates to zero.  
 Hence, the resulting bound on ${\cal I}_K$ is zero for fixed points of the deterministic dynamics (including the unstable fixed points).  It is positive for other values of $m$.



If the growth process has multiple stable states, this bound (obtained with control forces independent of time) does not provide an accurate estimate of ${\cal I}_K$.  The reason is similar to the  irreversible model.  To derive improved bounds we use the fact that the average action is very small if $b$ is close to $m_{\rm U}$.  We then follow the method that was illustrated in Fig.~\ref{fig:rateFn}.  We introduce the additional parameter $k^*$ and we construct trajectories that have magnetisation $b$ for $k<k^*$, but then follow the natural dynamics of the model for $k>k^*$.  The parameters $b,k^*$ are chosen such that the final magnetisation $m_K$ is close to its target value $m$.  The action can then be minimised by taking $b$ close to $\mU$ and $k^*/K$ to be small.

Compared to the irreversible model, the method for choosing $b,k^*$ is different; one must also fix $\lambda$, which determines $c'$ via (\ref{equ:cprime}).
Define $n_k = N_k/k$.  Following~\cite{Klymko2017}, 
we derive analogues of (\ref{equ:dmdu}), which are differential equations for the mean of $n_k$ and $m_k$ as a function of $u=\log k$.  These are
\beq
\frac{\mathrm{d}m}{\mathrm{d}u} = \frac{  (1-m^2) \sinh Jm - cm }{ n[ c + \cosh Jm - m\sinh Jm ]  }  
\label{equ:dmdu-rev}
\eeq  
and
\beq 
\frac{\mathrm{d}n}{\mathrm{d}u}  = -n + \frac{c - \cosh Jm + m\sinh Jm}{c + \cosh Jm - m\sinh Jm} \; .
\label{equ:dndu-rev}
\eeq
These equations will be solved forwards in time, in contrast to the irreversible model.  (The reason is that $m$ prescribes a target for $m_K$ but there is no target for $n_K$, so there are not enough boundary conditions to solve backwards in time.)  

Suppose that we are given some $K$ and some target $m$  for  $m_K$.  We propose an initial guess for $b$ that is between $\mU$ and $m$.  For this $b$, we minimise (\ref{equ:A-homog-rev2}) over $\lambda$ as above, to obtain a controlled process with $b$ as the steady state magnetisation.  This minimisation fixes the parameter $\lambda$ and hence also $c'$.
We then use the steady state of this controlled process as an initial condition and 
solve (\ref{equ:dmdu-rev},\ref{equ:dndu-rev}) forwards in $u$, starting from an (arbitrary) initial value $u_0$.  We stop the solution when the magnetisation hits the target.  This happens at some $u=u_1$ and we set $u_1=\log K$ so that the magnetisation will hit the target at the required time.  This requires that we identify $k^*=\log u_0$ as the point when we remove the control forces and allow the system to start evolving according to (\ref{equ:dmdu-rev},\ref{equ:dndu-rev}). Hence $k^*=K \ee^{u_0-u_1}$. 
Given the initial choice $b$, this yields a value for $k^*$ such that the system with time-dependent control forces will hit the target $m$.  The average action for this process is given by the product of (\ref{equ:A-homog-rev2}) and $k^*$, which is straightforward to evaluate.  

It remains to optimise the choice of $b$.  By analogy with (\ref{equ:sqrt},\ref{equ:A-sqrtK}), we choose this parameter such that the average action for the controlled process is
\beq
\langle {\cal A}/K\rangle_{\rm con} \approx \frac{a_A}{\sqrt{K}} \; ,
\label{equ:action-target}
\eeq
where $a_A$ is a parameter of order unity (we take $a_A=0.1$).  Given a target $m$ (with $|m|<\mS$) and a (sufficiently-large) value of $K$, it is possible to choose an initial guess for $b$ such that (i) the left hand side of (\ref{equ:action-target}) is larger than the right hand side, and (ii) the left hand side is reduced by moving $b$ closer to $\mU$.  Then, one may move $b$ towards $\mU$ in suitably-chosen steps until one finds parameters $(b,k^*)$ that solve (\ref{equ:action-target}).  The method for computation of $\langle {\cal A}\rangle_{\rm con}$ also fixes the values for $(\lambda,c')$, as described above.  

In this procedure, there is only one pitfall, which is  similar to the irreversible case: Eqs.~(\ref{equ:dmdu-rev},\ref{equ:dndu-rev}) are only applicable if the distribution of $m_{k^*}$ is sharply-peaked. (Specifically, its mean $b$ should differ from $\mU$ by an amount that is much larger than its standard deviation, which is of order $1/\sqrt{k^*}$).  This requirement is always satisfied if $K$ is large enough.  

Combining the ingredients, it follows that for large-$K$, we have established a bound on ${\cal I}_K$ that scales as
\beq
{\cal H}_K(m,\eps) \simeq \frac{a_A}{\sqrt{K}} \;.
\label{equ:HaA}
\eeq
This bound tends to zero at large $K$.
We emphasise that while this procedure for fixing $b,k^*$ is numerical, it does not involve any simulation of the growth process, only minimisation of (\ref{equ:A-homog-rev2}) and numerical solution of (\ref{equ:dmdu-rev},\ref{equ:dndu-rev}).

For cases where the model has two steady states, this method operates in the same way as the irreversible model and yields the same results.  (The unstable fixed point has $m_{\rm U}=0$ in this case.)  
By (\ref{equ:HaA}) one has ${\cal H}_K(m,\eps)\to0$ as $K\to\infty$ so the rate function in (\ref{equ:ldp-rev}) reduces to $I(m)=0$ for $|m| \leq \mS$.  All details of this computation are very similar to the irreversible model: for reasons of brevity we do not show numerical results in this case.

The more interesting situation occurs when the model has three steady states.  In this case we have performed simulations of the controlled process, in order to obtain bounds on ${\cal I}_K$.  Results are shown in Fig.~\ref{fig:rateFnRev}(b,c,d), following the procedure given above for determination of $(b,k^*,\lambda,c')$.  The trajectories of the controlled process remain close to the unstable fixed point for $k<k^*$, and diverge from it at later times.  Depending on the value of $b$, they may be attracted towards the fixed point at $0$, or the one at $\mS$.  In either case, the method yields results that are consistent with (\ref{equ:HaA}) and sufficient to establish that ${\cal I}_K(m,\eps)\to0$ as $K\to\infty$. Hence $I(m)=0$ whenever $|m|\leq \mS$, as in the irreversible case.

It would be interesting to investigate further the scaling of the probability density $\rho_K$ for these values of $m$, as in Sec.~\ref{sec:beyond-ldp}.  We anticipate similar results to that section, but a detailed analysis is beyond the scope of this work.

\section{Discussion}
\label{sec:conc}


These models of cluster growth show rich and interesting behaviour, both for typical trajectories~\cite{Klymko2017} and for large deviations~\cite{kggw18}.   They describe well-mixed clusters, in the sense that growth rates depend on the mean magnetisation and not, for example, the magnetisation near the boundary of the cluster.  This often results in a self-averaging property, so that fluctuations are small when clusters are large.  However, in cases where the deterministic dynamics has multiple fixed points (including unstable ones), large fluctuations are still possible, because trajectories may remain close to the unstable fixed point for large times, before eventually leaving it and converging (slowly) to a stable steady state.  See Fig.~\ref{fig:rateFn}(a) and Fig.~\ref{fig:rateFnRev}(b).

At the level of large deviations, these trajectories are associated with large fluctuations and manifest in a rate function $I(m)$ that is zero whenever $|m|<\mS$, recall Fig.~\ref{fig:rate-analytic}.  The probability to find a magnetisation in this range is not suppressed exponentially in $K$.  For the irreversible model, we have shown in Sec.~\ref{sec:beyond-ldp} that these probabilities decay as power laws in $K$ and we expect similar behaviour for the reversible model too.  


We have also emphasised that the models are not ergodic and do not fit the classes considered by~\cite{Lecomte2007,Chetrite2015}.  They can be expressed as non-Markovian processes and analysed using methods from~\cite{Harris2009,Harris2015}.  As noted in those works, this can lead to complex behaviour even for typical fluctuations, see Sec.~\ref{sec:harris-regime}.  

It is also useful to recall the analogy between large deviation theory and equilibrium thermodynamics~\cite{Lecomte2007,Garrahan2009,Touchette2009}.  Within this framework, probabilities of individual trajectories (in $d$ dimensions of space and 1 dimension of time) are analogous to configurations of equilibrium systems in spatial $d+1$ dimensions.  The growth models have no spatial degrees of freedom so $d=0$; this means that trajectories of the growth model are analogous to configurations of a one-dimensional Ising model, where $s_k$ is interpreted as the $k$th spin in the chain.  The energy of a configuration of this Ising model is 
\beq
E(\bm{s})= E_0 - \sum_{k=1}^K \log p_k(s_k|m_{k-1})
\label{equ:en-ising}
\eeq 
with $\bm{s}=(s_1,s_2,\dots,s_K)$ and $m_{k-1}$ given by (\ref{equ:def-m}); also $E_0$ is an arbitrary additive constant and the analogy requires that the temperature $T=1$.   Equ.~(\ref{equ:en-ising}) corresponds to an Ising model with long-ranged interactions, while the standard Markovian class of models would have only nearest-neighbour interactions.  

Within this analogy, the rate function in the LDP corresponds to a free energy in the equilibrium system.  This means that the results of Fig.~\ref{fig:rate-analytic} and Fig.~\ref{fig:rateFnRev}(a) somewhat resemble a double-tangent construction, which would usually be associated with phase coexistence (it is equivalent to the Maxwell construction, see Sec.~4.7 of~\cite{Huang-intro}). The analogy between dynamical large deviations and phase coexistence is discussed, for example, in~\cite{Hedges2009,Elmatad2010,Nemoto2017first}.  The physical analogue of phase coexistence in dynamical trajectories may depend on system details, but one possibility is that trajectories that realise the rare event of interest show different behaviour in early-time and late-time regimes, as in Fig.~2A of~\cite{Elmatad2010} and Fig.~4B of~\cite{Hedges2009}.%

In the growth models considered here, the analogy with equilibrium phase coexistence is not complete because of the long-ranged interactions in the Ising energy (\ref{equ:en-ising}).  From Fig.~\ref{fig:rateFn}(a), one sees that the trajectories that realise the relevant rare events have qualitatively different behaviour in the early-time regime ($k<k^*$) and the late-time regime ($k>k^*$), similar to the behaviour for Markovian models~\cite{Hedges2009,Elmatad2010}.  
However the behaviour in the late-time regime is not at all stationary, for example the typical magnetisation $m_k$ depends on $k$ throughout the range $k^*<k<K$.  This is contrary to the behaviour in Markovian models~\cite{Hedges2009,Elmatad2010,Nemoto2017first} where averages depend weakly on time \emph{within} the late-time and early-time regimes, even if they differ strongly \emph{between} these two regimes.  For this reason, we prefer not to use the terminology of phases and phase coexistence to describe the behaviour shown in Fig.~\ref{fig:rate-analytic}.  Nevertheless, the behaviour of the rate function is the same as one would obtain from a double-tangent construction.  

We also emphasise that while the rate functions for these growth models are never concave, the double-tangent construction does not hold generally in non-Markovian systems~\cite{Duffy2008} nor even in Markovian systems on non-compact state spaces~\cite{Nickelsen2018}  -- in such cases, the applicability of the double-tangent construction has to be tested on a case-by-case basis.  This is similar to analysis of thermodynamic phase coexistence in systems with long-ranged interactions, where the applicability of the Maxwell construction depends on the decay of the interaction potential~\cite{campa2009}.  For the growth models considered here, we have shown that the construction is applicable.

Finally, we comment on the usefulness of the bound (\ref{equ:IK-bound}) for numerical estimation of small probabilities, as in~\cite{kggw18,Whitelam2018sampling}.  
Our results here confirm that suitable choices of the controlled dynamics can make this bound accurate (see for example Figs.~\ref{fig:bound-good}b and \ref{fig:J-16-peak}).  However, construction of the relevant controlled dynamics in those cases required  detailed understanding of the dynamical behaviour of the model (including analytical estimates of the action).   Our conclusion is that this method is only reliable if one already has a precise understanding of the mechanism by which the relevant rare events (large deviations) will occur.    In this case, one may design a controlled process with this mechanism in mind.  However, experience with a range of model systems (see for example Sec.~3.4 of~\cite{Jack2015b}) indicates that it is difficult to predict suitable controlled dynamics, without prior theoretical analysis.  If one evaluates the bound (\ref{equ:IK-bound}) using a controlled process does not fully account for the mechanism of the rare event, one may expect to obtain bounds that are not accurate estimates of the probabilities of interest.%

\begin{acknowledgments}
I would like to thank Steve Whitelam for many interesting discussions about growth models and numerical sampling methods.  I am also grateful to Rosemary Harris, Hugo Touchette, and Juan P. Garrahan for useful discussions, including those related to LDPs in non-Markovian processes, and the possibility of optimal control forces that depend explicitly on time.
\end{acknowledgments}

\vfill

\bibliographystyle{apsrev4-1}
\bibliography{dev}

\end{document}